\documentclass[aps,showpacs,preprintnumbers,amsmath,amssymb]{revtex4}

\usepackage{float}
\usepackage{graphics,epsfig}
\usepackage{graphicx}
\usepackage{dcolumn}
\usepackage{bm}

\begin{document}
\baselineskip=0.8 cm
\title{{\bf Quasinormal modes of charged squashed Kaluza-Klein black
holes in the G\"{o}del Universe}}
\author{Xi He}\affiliation{Department of Physics, Fudan
University, 200433 Shanghai, China}

\author{Songbai Chen}
\email{csb3752@163.com} \affiliation{Institute of Physics and
Department of Physics, Hunan Normal University,  Changsha, 410081
Hunan, China}

\author{Bin Wang}
\email{wangb@fudan.edu.cn} \affiliation{Department of Physics,
Fudan University, 200433 Shanghai, China}

\vspace*{0.2cm}
\begin{abstract}
\baselineskip=0.6 cm
\begin{center}
{\bf Abstract}
\end{center}
We study the quasinormal modes of scalar perturbation in the
background of five-dimensional charged Kaluza-Klein black holes
with squashed horizons immersed in the G\"{o}del universe. Besides
the influence due to the compactness of the extra dimension, we
disclose the cosmological rotational effect in the wave dynamics.
The wave behavior affected by the G\"{o}del parameter provides an
interesting insight into the G\"{o}del universe.

\end{abstract}

\pacs{ 04.70.Dy, 95.30.Sf, 97.60.Lf } \maketitle
\newpage

\section{Introduction  }

Quasinormal mode (QNM) plays an essential role in the study of black
hole physics. It is known as a unique fingerprint of the black hole
existence, since quasinormal frequencies only depend on black hole
parameters. It is expected that QNM can be detected in the
gravitational wave observations to be realized in the near future
\cite{1}. Besides its potential observational interest in astronomy
to detect the black hole existence, theoretically QNM is believed as
a testing ground of fundamental physics. It is widely believed that
the study of QNM can help us get deeper insights into the
AdS/CFT\cite{3} and dS/CFT\cite{4} correspondences (for a review on
this topic and more complete list of references can be found in
\cite{5}\cite{add}). Recently further motivation of studying the
QNMs has been pointed out in \cite{6,7,8} by arguing that QNMs can
provide a phenomenological signature of black hole thermal phase
transition. Further evidences of the non-trivial relation between
the dynamical and thermodynamical properties of black holes were
provided in\cite{9,10}. Furthermore string theory has the radial
prediction that spacetime has extra dimensions and gravity
propagates in higher dimensions, QNM has recently been argued as a
useful tool to disclose the existence of the extra dimensions
\cite{11,12,13}. Most studies of the QNM are concentrated on black
holes immersed in the rather idealized isotropic homogeneous
universe.

It was argued that it is more reasonable to consider the universe
background as homogeneous but with global rotation\cite{14}. An
original exact solution of Einstein equations with pressureless
matter and negative cosmological constant for the rotating universe
was found by G\"{o}del \cite{15} in four-dimensional spacetime,
which exhibits close timelike curves through every point.
Supersymmetric generalizations of the G\"{o}del universe in five
dimensions have been found in \cite{16}. Exact solutions describing
various black holes embedded in a G\"{o}del universe have been
presented in \cite{17,18}. Applying the ¡®squashing¡¯
transformation, the new ¡®squashed¡¯ Kaluza-Klein(KK) black hole
solutions in the rotating universe have been obtained in
\cite{19,20}. In \cite{21,22} the possible influence of the rotation
of the universe on the QNM of the Schwarzschild-G\"{o}del black hole
and the Hawking radiation in the rotating G\"{o}del black hole have
been investigated respectively. It was found that the quasinormal
spectrum and Hawking radiation are considerably affected by the
rotating cosmological background. In this work we are going to
explore specifically the QNM of the new solution obtained by
applying the squashing transformation to the general charged black
hole embedded in the G\"{o}del universe \cite{19}. Besides the
influence due to the compactness of the extra dimension on the
QNM\cite{10,23}, here we will also disclose the cosmological
rotation effect. As done in \cite{21,22} we will limit our attention
to the case of slow rotation of the cosmological background. This is
mainly because that phenomenologically the small rotation of the
universe is the most reasonable situation. In addition, the small
rotation can help to separate the variables in perturbation
equations.

The plan of the paper is as follows. In Sec. II we first go over the
charged squashed KK G\"{o}del black hole background and then derive
the master equation of scalar perturbation in the limit of small
G\"{o}del parameter $j$. In Sec. III, we discuss our numerical
result on the quasinormal frequencies for the scalar field
perturbations. In the end  we present our conclusions and further
discussions.

\section{scalar field perturbation of the charged squashed KK black hole in G\"{o}del Universe  }

The line element of the general charged rotating squashed KK
G\"{o}del black hole is\cite{19}
\begin{eqnarray}\label{1}
ds^2=-f(r)dt^2+{k(r)^2\over V(r)}dr^2-2g(r)\sigma_3 dt +h(r)
\sigma_3^2+{r^2\over
4}\bigg[k(r)\big(\sigma_1^2+\sigma_2^2\big)+\sigma_3^2\bigg],
\end{eqnarray}
where
\begin{eqnarray}
&&\sigma_1=cos\psi d\theta+sin\psi sin\theta d\phi,\\ \nonumber
&&\sigma_2=-sin\psi d\theta+cos\psi sin\theta d\phi, \\ \nonumber
&&\sigma_3=d\psi+cos\theta d\phi,
\end{eqnarray}
with coordinates $\theta \in [0,~\pi)$, $\phi \in [0, ~2 \pi)$ and
$\psi \in [0, ~4\pi)$, and $r$ runs in the range $(0,~ r_\infty)$.
The metric functions read
\begin{eqnarray}
f(r)&=&1-{2M\over r^2}+{q^2\over r^4},\\ \nonumber g(r)&=&jr^2
+3jq+{(2M-q)\tilde{a} \over 2r^2}-{q^2 \tilde{a}\over 2r^4},\\
\nonumber h(r)&=&-j^2
r^2(r^2 +2M+6q)+3jq\tilde{a} +{(M-q)\tilde{a}^2\over 2r^2}-{q^2 \tilde{a}^2 \over 4r^4}\\
\nonumber V(r)&=& 1-{2M\over r^2} +{8j (M+q)\big
[\tilde{a}+2j(M+2q)\big]\over r^2}
\\ \nonumber && +{2(M-q)\tilde{a}^2 +q^2 \big[1-16 j \tilde{a} -8j^2 (M+3q)\big]\over r^4},
\\ \nonumber k(r)&=&{V(r_\infty)r_\infty^4 \over (r^2 -r_\infty
^2)^2}.
\end{eqnarray}

Constant $M$, $q$, $\tilde{a}$ represent the mass, charge and
rotation of the black hole, and $j$ denotes the scale of the
G\"{o}del background. We will concentrate on the black hole without
rotation by setting $\tilde{a}=0$.  When $j=0, \tilde{a}=0$,  Eq.
(\ref{1}) reduces to the five-dimensional charged KK black hole with
squashed horizon\cite{24}. When $r_\infty \rightarrow \infty$
($k(r)\rightarrow 1$), the squashing effect disappears and one
recovers the five-dimensional charged black hole in the G\"{o}del
universe.

Considering the reasonable phenomenology that our universe must
rotate slowly even if there is global rotation in the cosmological
background, we will expand Eq.(\ref{1}) in the small $j$ limit by
discarding terms over the order $O(j^2)$. In \cite{21,22}, it was
argued that small $j$ limit is required to separate variables in
perturbation equations. Using coordinate transformation
$\rho=\rho_0 {r^2\over r_\infty^2 -r^2}$ and
$\tau=\sqrt{V(r_\infty)} t$, Eq.(\ref{1}) can be rewritten in the
form
\begin{eqnarray}
ds^2=-F(\rho)d\tau^2 +{K^2(\rho)\over F(\rho)}d\rho^2 +\rho^2
K^2(\rho) \big(d\theta^2+sin^2\theta d\phi^2 \big)+{r_\infty^2 \over
4K^2(\rho)}\sigma_3^2- 2 H(\rho)\sigma_3 d\tau,
\end{eqnarray}
with
\begin{eqnarray}
&&F(\rho)=\bigg(1-{\rho_+ \over \rho}\bigg)\bigg(1-{\rho_- \over
\rho}\bigg),\\ \nonumber &&K(\rho)^2=1+{\rho_0 \over \rho},\\
\nonumber &&H(\rho)={r_\infty^3 \over 2 \rho_0 K^2}j+{3 r_\infty^2
\over \rho_0}\sqrt{\rho_+ \rho_-}j,
\end{eqnarray}
where $\rho_0^2={r_\infty^2\over 4}V(r_\infty)$, so that
$r_\infty^2=4(\rho_++\rho_0)(\rho_-+\rho_0)$. $\rho_+, \rho_-$
denote the outer and inner horizons of the black hole in the new
coordinate. For simplicity we will follow \cite{10} to take
$\rho_+=1-{a\over 2}$ and $\rho_-={a\over 2}$ and adjust the
parameter $a$ to indicate the extent on how far the outer horizon is
away from the inner horizon.

The scalar field perturbations in charged squashed KK G\"{o}del
black hole background are governed by the Klein-Gordon equation
$\Box \Phi =0$. We can separate the variables of the field equation
by adopting the limit of small $j$ and representing the wave
function as $\Phi(\tau, \rho, \theta, \phi,\psi)=e^{i \omega \tau +
im \phi -i \lambda \psi} R(\rho)S(\theta)$. Then the perturbation
equations reduce to
\begin{eqnarray}
&&{F\over \rho^2 K^2}{\partial \over \partial \rho}\bigg[\rho^2 F
{\partial R \over \partial \rho}\bigg]+\bigg[\omega^2-{8\omega
\lambda H K^2 \over r_\infty^2}-{4F\lambda^2 K^2 \over
r_\infty^2}-{F\over \rho^2 K^2}E_{lm\lambda}\bigg]R=0
\label{Radial}, \\ && {d^2S\over d\theta^2}+cot\theta {dS \over
d\theta} +\bigg[E_{lm\lambda}-(m-\lambda cos\theta)^2
csc^2\theta\bigg]S=0\label{spherical},
\end{eqnarray}
where $E_{lm\lambda}=l(l+1)-\lambda^2$ is the eigenvalue of the
spin-weighted spherical function (\ref{spherical}), $\lambda$ is
the separation constant of the fifth dimension.

Boundary conditions on the wave function $R(\rho)$ at the outer
horizon and the spatial infinity can be expressed as
\begin{eqnarray}
R(\rho)\sim \left\{
  \begin{array}{ll}
    {(\rho-\rho_+)^{\alpha}},
&{\hbox{$\rho$ $\rightarrow$ $\rho_+$};}
\\ \\
    {\rho^{\gamma}
e^{i\chi\rho}},~~~~ & \hbox{$\rho\rightarrow\infty$.}\label{bd1}
  \end{array}
\right.
\end{eqnarray}
A solution of equation (\ref{Radial}) that satisfies  the above
boundary condition can be written as
\begin{eqnarray}
R(\rho)=e^{i(\rho-\rho_-)\chi} (\rho-\rho_+)^{\alpha}
(\rho-\rho_-)^{\beta+\gamma} \sum_{m=0}^{\infty} a_m
\bigg({\rho-\rho_+ \over \rho- \rho_-}\bigg)^m,
\end{eqnarray}
where constants $\alpha$, $\beta$, $\gamma$ and $\chi$ are
\begin{eqnarray}
\alpha &=&     {1\over \sqrt{\rho_0}  \big(\rho_- - \rho_+\big)
}\bigg\{-\rho_+^2 \big(\rho_0 + \rho_+\big) \omega \\
\nonumber &&  \bigg[-8 j \rho_+ \bigg(3 \sqrt{\rho_- \rho_+}+
{r_\infty\over 2}\bigg)\lambda + \rho_0 \bigg(\rho_+ \omega - 24 j
\sqrt{\rho_- \rho_+}
\lambda\bigg)\bigg]\bigg\}^{1\over 2}, \\
\beta&=&{1\over \sqrt{{\rho_0}}
   \big({\rho_+}-{\rho_-}\big)} \bigg\{-{\rho_-}^2 \big({\rho_0}+{\rho_-}\big) \omega \\ \nonumber &&\bigg[{\rho_0}
   \bigg({\rho_-} \omega-24 j \sqrt{{\rho_-} {\rho_+}} \lambda \bigg)-8 j
   {\rho_-} \left(3 \sqrt{{\rho_-
   } {\rho_+}}+{r_\infty\over 2}\right) \lambda \bigg]\bigg\}^{1\over 2}, \\
\gamma&=& {-i\over {{\rho_0} {r_\infty^2 \over 2} \sqrt{
   \omega^2-{{4{\lambda}^2\over{r_\infty^2} }-
   {8 j \over \rho_0}\big(3 \sqrt{{\rho_-}
   {\rho_+}}+{r_\infty \over 2}\big)
   \omega {\lambda}}}}}   \bigg\{{\rho_0} (2 {\rho_0}+{\rho_-}+{\rho_+})
\lambda^2 \\ \nonumber && +2 j r_\infty^2 \bigg[2
({\rho_-}+{\rho_+})
   \bigg(3 \sqrt{{\rho_-}{\rho_+}}+{r_\infty \over 2}\bigg)+{\rho_0} \bigg(6 \sqrt{ {\rho_-}
   {\rho_+}}+{r_\infty \over 2}\bigg)\bigg] \omega {\lambda}\\ \nonumber &&-{\rho_0} {r_\infty^2 \over 4}
  \bigg[ ({\rho_0} +2 {\rho_-}+2 {\rho_+} )\omega^2+2 i \sqrt{
   \omega^2-{{
   { 4\lambda}^2\over r_\infty^2}-{8 j\over \rho_0}\bigg(3 \sqrt{{\rho_-}
   {\rho_+}}+{r_\infty \over 2}\bigg) \omega \lambda} } \bigg]\bigg\}, \\
\chi &=& \sqrt{\omega^2-\frac{8 j \left(3 \sqrt{{\rho_-}
   {\rho_+}}+{r_\infty\over 2}\right)
   \lambda  \omega}{{\rho_0}}-\frac{4\lambda ^2}{{r_\infty^2}}}.
\end{eqnarray}

We will employ the continued fraction method to find the accurate
quasinormal frequencies of the charged squashed KK black hole in
G\"{o}del universe. The radial equation (\ref{Radial}) for the
perturbations can be solved using a series expansion around some
irregular points, such as the event horizon. The coefficients $a_m$
of the expansion are then determined by a recursion relation
starting from $a_0=1$ with the form
 \begin{eqnarray}\label{se}
&&\alpha_0 a_1 +\beta_0 a_0=0, \\ \nonumber && \alpha_m
a_{m+1}+\beta_m a_m + \gamma_m a_{m-1}=0, ~~~m=1,~2,...
 \end{eqnarray}
where the recursion coefficients $\alpha_m$, $\beta_m$, $\gamma_m$
are given by
\begin{eqnarray}
&&\alpha_m=m^2+(C_0+1)m+C_0, \\ \nonumber
&&\beta_m=-2m^2+(C_1+2)m+C_3, \\ \nonumber
&&\gamma_m=m^2+(C_2-3)m+C_4-C_2+2.
\end{eqnarray}
$C_m$s are functions of the frequency $\omega$ and when $j=0$,
they reduce to (23)-(26) in \cite{10}. For conciseness, here we
will not write out their tedious expressions when $j\ne 0$. The
boundary conditions are satisfied when the continued fraction
condition on the recursion coefficients holds. The series in
(\ref{se}) converge for the given $l$. The frequency $\omega$ is a
root of the continued fraction equation
\begin{eqnarray}\label{continue}
\bigg[\beta_m-{\alpha_{m-1}\gamma_m\over
\beta_{m-1}-}{\alpha_{m-2}\gamma_{m-1}\over
\beta_{m-2}-}...{\alpha_0 \gamma_0 \over
\beta_0}\bigg]=\bigg[{\alpha_m \gamma_{m+1}\over
\beta_{m+1}-}{\alpha_{m+1}\gamma_{m+2}\over
\beta_{m+2}-}{\alpha_{m+2}\gamma_{m+3}\over
\beta_{m+3}-}....\bigg],\;\;\;(m=1,2,...).
\end{eqnarray}
Solving the above continued fraction equation (\ref{continue})
numerically, we can obtain the QNMs for the charged squashed KK
black hole in G\"{o}del universe.

\section{The behavior of QNMs for the charged squashed KK black hole in G\"{o}del Universe}

\begin{figure}[h]
\includegraphics[width=10cm]{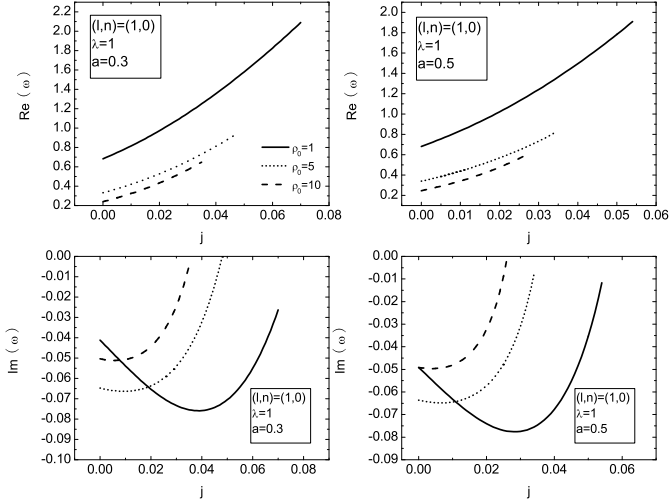}
\caption{\label{varyj}The behaviors of $Re(\omega)$ and
$|Im(\omega)|$ with the change of $j$ for chosen $\rho_0$. The
solid line is for $\rho_0=1$, the dotted line for $\rho_0=5$ and
the dashed line  for $\rho_0=10$. In plotting the figure, we have
chosen $\lambda=1$, $(l, n)=(1, 0)$. }
\end{figure}

\begin{table}
\caption{\label{Table}Quasnormal frequencies of scalar perturbations
of charged squashed Kaluza-Klein black holes in the G\"{o}del
universe with the change of j for chosen $\rho_0$, for $l=1$, $n=0$
and $a=0.3$. }
\begin{center}
\begin{tabular}{|c||c|c|c|}
\hline
$j$ & $\rho_0=30 $& $\rho_0=60$& $\rho_0=80$\\
\hline
$0.001$&$0.148491-0.0310149i$&$0.107443-0.0222711i $&$0.094073-0.0193677i$\\
$0.002$&$0.155686-0.0311394i$&$0.114733-0.0223486i $&$0.101447-0.0194209i$\\
$0.003$&$0.163248-0.0311927i$&$0.122536-0.0223191i $&$0.109415-0.0193456i$\\
$0.004$&$0.171186-0.0311677i$&$0.130872-0.0221666i $&$0.118004-0.0191193i$\\
$0.005$&$0.179511-0.0310563i$&$0.139760-0.0218721i $&$0.127240-0.0187139i$\\
\hline \hline
$j$ & $\rho_0=100 $& $\rho_0=300$& $\rho_0=350$\\
\hline
$0.001$&$0.084934-0.0173686i$&$0.052190-0.0101018i $&$0.048884-0.0093552i$\\
$0.002$&$0.092390-0.0173999i$&$0.060319-0.0099664i $&$0.057152-0.0091842i$\\
$0.003$&$0.100515-0.0172826i$&$0.069674-0.0094924i $&$0.066758-0.0086265i$\\
$0.004$&$0.109341-0.0169864i$&$0.080350-0.0085354i $&$0.077817-0.0074964i$\\
$0.005$&$0.118899-0.0164730i$&$0.092448-0.0068664i $&$0.090443-0.0054893i$\\
\hline
\end{tabular}
\end{center}
\end{table}

We now report our numerical results of QNMs for the charged
squashed KK black hole in G\"{o}del universe. The fundamental QNMs
will be our primary interest, since fundamental modes dominate in
the late time oscillations. We generalized our previous numerical
program used in \cite{10} by including small G\"{o}del parameter
$j$. When $j=0$, it was shown in \cite{10} that our numerical
calculation is reliable and precise by comparing with results in
\cite{23}.

In Fig.(\ref{varyj}) we display the dependence of quasinormal
frequencies of scalar perturbations on the G\"{o}del parameter $j$
in the regime when  $j$ is small. We found that for chosen values
of $(l, n)$, $\lambda=1$, $a$ and $\rho_0$, the real part of
quasinormal frequency increases with the increase of $j$. Our
numerical results are for small values of $\rho_0$, which means
that the squashed effect is strong in our result. The behavior of
the oscillation frequency is different from that found in the
Schwarzschild black hole in the G\"{o}del universe\cite{21}, where
it was observed that the oscillation frequency linearly decreases
with $j$. Considering that the quasinormal spectrum is sensitive
to boundary conditions at the event horizon and at spatial
infinity and in our case the asymptotic structures near the event
horizon and the spatial infinity are different from those in the
background of \cite{21}, it is not strange to find different
behaviors of the real part frequencies in these two different
backgrounds. The spectrum we observed is considerably affected by
the squashed effect.

The behavior of the imaginary part of the quasinormal frequency on
the G\"{o}del scale parameter $j$ has also been shown in Fig.
(\ref{varyj}). For not very small $\rho_0$, we observed that the
absolute values of the imaginary frequency monotonically decrease
with the increase of $j$. This tells us that when the universe has
more rotation, it is more difficult for the perturbation around the
black hole to calm down. The scalar field damps more slowly for
bigger values of $j$. The dependence of damping rate on the
G\"{o}del parameter $j$ agrees with that found in \cite{21}. In the
limit of small $j$ the QNMs govern the decay of the scalar field at
late times. For severely squashed case with very small $\rho_0$, we
found that with the increase of $j$, the absolute value of the
imaginary frequency first increases and then decreases. More
detailed numerical analysis on this phenomenon has been carried out
and the results are shown in table \ref{Table} for $a=0.3$. It is
clear to see that with the increase of $\rho_0$, the maximum value
of the absolute imaginary frequency appears for smaller $j$ and when
$\rho_0 \geq 300$, the imaginary frequency changes monotonically
with the increase of $j$. For bigger $a$, the critical value to
obtain the monotonic behavior of the imaginary frequency becomes
smaller. This shows that when the squashed effect is considerable,
the perturbation around the black hole will even decay faster in the
slower rotating universe. This can be attributed to the fact that
near the event horizon the black hole is severely squashed and the
asymptotic structure there has been drastically changed. However
with the increase of $j$, the squashed effect seems diluted and the
damping time increases again with $j$.

\begin{figure}[h]
\includegraphics[width=10cm]{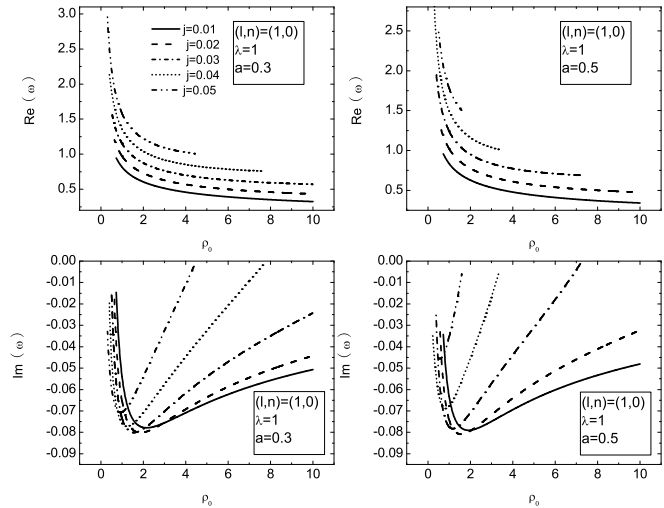}
\caption{\label{varyrho0} The behaviors of $Re (\omega)$ and $Im
(\omega)$ with the change of $\rho_0$ for chosen $j$. The solid
line is for $j=0.01$, the dashed line for $j=0.02$, dashed-dotted
line for $j=0.03$, short dotted line for $j=0.04$ and the left
line is for $j=0.05$. }
\end{figure}

In Fig. (\ref{varyrho0}) we show the dependence of quasinormal
frequencies of scalar perturbations around the charged squashed KK
black holes in G\"{o}del universe on the parameter $\rho_0$ for
different chosen values of $j$. For fixed parameters of $(l, n),
\lambda, a$ and chosen $j$, we see that the real part of the
quasinormal frequency decreases with the increase of $\rho_0$,
which is consistent with the result in the non-rotating squashed
KK black holes when $j=0$ observed in \cite{10,23}.

\begin{figure}[h]
\includegraphics[width=10cm]{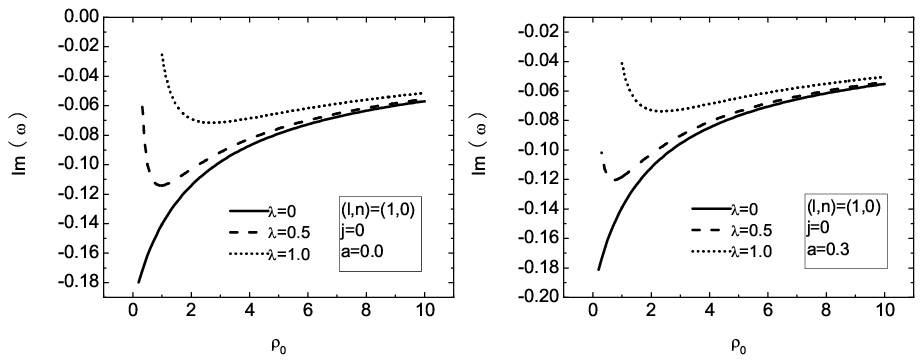}
\includegraphics[width=10cm]{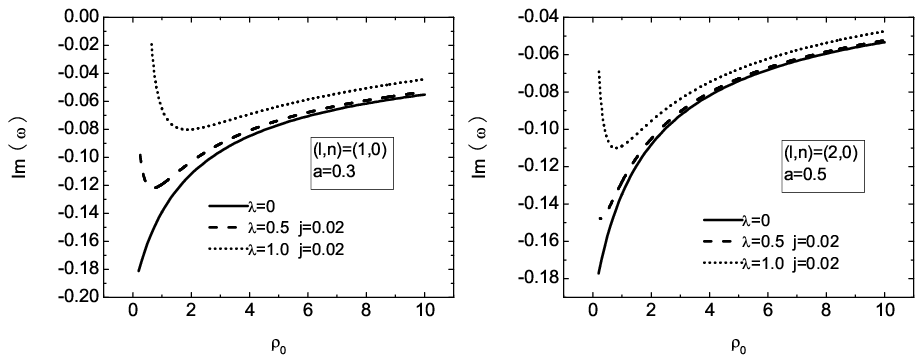}
\caption{\label{varyLa}Graphs of $Im(\omega)$ with the change of
$\rho_0$ for different $\lambda$. The solid line is for
$\lambda=0$, the dashed line is for $\lambda=0.5$ and the dotted
line for $\lambda=1$. }
\end{figure}

When $\lambda\ne 0$, we see from Fig. (\ref{varyrho0}) that the
absolute imaginary part of the quasinormal frequency first
increases and then decreases with the increase of $\rho_0$. This
result holds for all selected small $j$ and consistent with the
findings in the gravitational QNMs in \cite{23} when $K=1,2$ and
$j=0$ there. For bigger values of $j$, we observed deeper U-turn
in the absolute imaginary frequency and it is easier for
$|\omega_I|\rightarrow 0$ if the black hole is not severely
squashed. This is consistent with that shown in Fig.
(\ref{varyj}), telling us that the damping time of the
perturbation is longer if the universe rotates more.  We have also
investigated the imaginary part of the quasinormal frequency when
$\lambda=0$, the result is shown in Fig.(\ref{varyLa}). Comparing
with the case when $\lambda\ne 0$, we found that there is no
turning point in $|\omega_I|$ and $|\omega_I|$ decreases
monotonically with the increase of $\rho_0$. This result holds for
all our selected values of $j$ and when $j=0$ it is consistent
with the gravitational perturbation result when $K=0$ in
\cite{23}. Remembering that $\lambda$ is the separation constant
due to the fifth dimension, we learnt that the turning point in
$|\omega_I|$ is brought by the extra dimensional effect. When
$j=0$, $|\omega_I|$ approaches to a constant nonzero value which
reduces to the result obtained in \cite{10,23}.

\begin{figure}
\includegraphics[width=12cm]{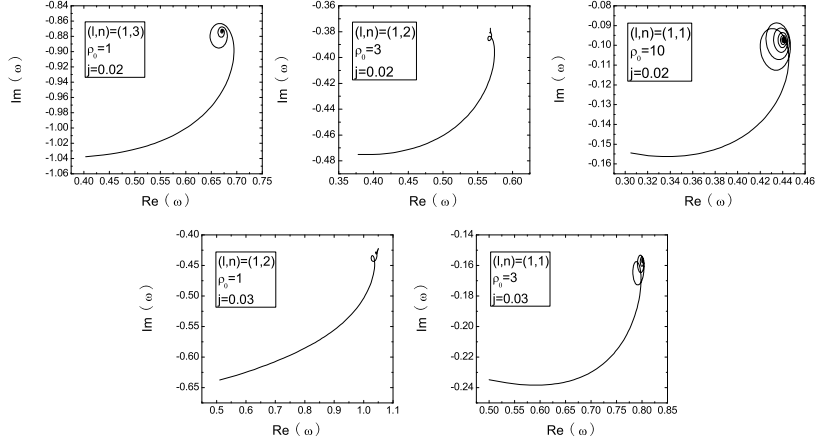}
\caption{\label{varya}The first spiral-like shape with the increase
of $\rho_-$ for $j=0.02, 0.03$ in the complex $\omega$ plane of
scalar QNMs around squashed charged KK black holes in G\"{o}del
universe. In the figure we have set $\lambda=1$, $l=1$ and
$\rho_0=1$, $3$ and $10$.}
\end{figure}

In addition to the fundamental modes, we have also calculated the
contribution of the overtones. The numerical behavior of the first
few overtones of the charged squashed KK black hole in G\"{o}del
universe is shown in Fig. (\ref{varya}). The trajectories described
by the modes in the complex-$\omega$ plan exhibit the spiral-like
behavior when $\rho_+$ approaches to $\rho_-$. For $j=0$, this
behavior was observed in asymptotically flat black holes in
\cite{9,25} and also in the squashed KK black holes in \cite{10}. We
observed that such a spiralling behavior sets in for larger overtone
number $n_c$ as the decrease of $\rho_0$ for selected $j$. For
$j=0$, this behavior was obtained in \cite{10}. For fixed $\rho_0$,
we found that with the increase of $j$, the spiralling behavior sets
in for smaller values of the overtone $n_c$.

\section{Conclusion and discussions}
We have investigated the scalar perturbation in the background of
charged squashed KK black holes immersed in a rotating
cosmological background. In the limit of small G\"{o}del parameter
$j$, the quasinormal frequencies have been calculated. It was
found that due to the squashed effect, the real part of QNM
frequency increases when the cosmological background rotates. For
the imaginary part of the quasinormal frequency, the strong
squashed effect will cause $|\omega_I|$ to increase with $j$ when
the universe does not rotate so fast, however when $j$ gets
bigger, $|\omega_I|$ will finally decrease with $j$. In the small
G\"{o}del parameter limit, all found modes are damping which shows
the stability of the charged squashed black hole in the G\"{o}del
universe against scalar perturbation. However we saw the tendency
that with the increase of $j$, especially for bigger $\rho_0$, the
$|\omega_I|$ tends zero, which tells us that the damping time
becomes longer and longer. This could imply that there is a danger
to have unstable charged squashed KK black hole in the G\"{o}del
cosmological background for big G\"{o}del parameter $j$. This
needs careful examination in the future. For big $j$ we have
problems in separating radial equations. On the other hand,
remembering that the stability of the spacetime is mainly
determined by the gravitational perturbation, it is of great
interest to generalize the study here to the gravitational
perturbations in the future.

In the homogeneous isotropic universe with $j = 0$, it was
observed that the complex $\omega$ plan starts to exhibit the
spiral-like shape when the black hole parameter reaches a critical
value and the nontrivial relation between the dynamical and
thermodynamical properties of black holes were claimed existing in
the loop behavior in the complex $\omega$ plan[8, 9]. In the
G\"{o}del universe with small $j$, although the spiral-like
behavior in the complex $\omega$ has also been observed, it is not
clear at this moment whether it can tell us some relation between
the thermodynamical and dynamical properties. For bigger $j$ in
our limit, the computation time becomes longer and especially when
two black hole horizons come closer. Careful examinations of the
dynamical properties through the QNMs and comparisons with the
thermodynamical properties in \cite{26} are called for for the
G\"{o}del universe.

\end{document}